\documentclass{llncs}
\usepackage{graphicx,epsfig,color}
\usepackage[shortend,ruled]{algorithm2e}
\usepackage{ifthen,verbatim,array,multirow}
\usepackage{latexsym,amssymb,amsmath,amsthm}
\usepackage[english]{babel}
\usepackage[utf8]{inputenc}

\usepackage{xspace}
\usepackage{tikz}
\usetikzlibrary{arrows,decorations,positioning,shapes}
\usepackage{setspace}
\usepackage{enumitem}
\setlist{nolistsep}
\usepackage{listings}

\usepackage{textcomp}
\usepackage{stmaryrd}
\usepackage{pifont}
\usepackage{soul}

\usepackage{booktabs}   
\usepackage{multirow}   
\usepackage{cite}       
\usepackage{subcaption} 
\usepackage{caption}
\usepackage{todonotes}  
\usepackage{tablefootnote}
\usepackage[most]{tcolorbox}

\let\llncssubparagraph\subparagraph
\let\subparagraph\paragraph
\usepackage{titlesec}
\let\subparagraph\llncssubparagraph

\usepackage[pdftex%
,colorlinks=true       
,linkcolor=midblue        
,citecolor=midblue        
,filecolor=black       
,urlcolor=midblue         
,linktoc=page           
,plainpages=false]{hyperref}
\addto\extrasenglish{%

}

\fontfamily{ptm}

\setlist{nolistsep}
\setitemize{noitemsep,topsep=3pt,parsep=2pt,partopsep=2pt}
\setenumerate{noitemsep,topsep=3pt,parsep=2pt,partopsep=2pt}

\titlespacing{\section}{0pt}{*2.5}{*1.0}
\titlespacing{\subsection}{0pt}{*1.5}{*0.75}
\titlespacing{\subsubsection}{0pt}{*1.0}{*0.5}

\makeatletter
\renewcommand\paragraph{\@startsection{paragraph}{4}{\z@}%
                                    {1.0ex \@plus0.35ex \@minus.15ex}%
                                    {-1em}%
                                    {\normalfont\normalsize\bfseries}}

\makeatother

\makeatletter
\def\thm@space@setup{%
  \thm@preskip=2pt \thm@postskip=2pt
}
\makeatother

\makeatletter
\renewenvironment{proof}[1][\proofname]{\par
  \vspace{-0.25\topsep}
  \pushQED{\qed}%
  \normalfont
  \topsep0pt \partopsep0pt 
  \trivlist
  \item[\hskip\labelsep
        \itshape
    #1\@addpunct{.}]\ignorespaces
}{%
  \popQED\endtrivlist\@endpefalse
  \addvspace{1pt plus 1pt} 
}
\makeatother

\titleformat{\paragraph}[runin]
            {\normalfont\normalsize\bfseries}{\theparagraph}{0.4em}{}

\newtheorem{nproposition}{Proposition}

\newtheorem{ndefinition}{Definition}

\newtheorem{nreduction}{Reduction}

\theoremstyle{remark}

\newtheorem{nremark}{Remark}

\theoremstyle{plain}

\newcommand{\vars}{\mathsf{vars}} 

\newcommand{\fml}[1]{{\mathcal{#1}}}

\definecolor{gray}{rgb}{.4,.4,.4}
\definecolor{midgrey}{rgb}{0.5,0.5,0.5}
\definecolor{middarkgrey}{rgb}{0.35,0.35,0.35}
\definecolor{darkgrey}{rgb}{0.3,0.3,0.3}
\definecolor{darkred}{rgb}{0.7,0.1,0.1}
\definecolor{midblue}{rgb}{0.2,0.2,0.7}
\definecolor{darkblue}{rgb}{0.1,0.1,0.5}
\definecolor{defseagreen}{cmyk}{0.69,0,0.50,0}

\newcommand{\jnoteF}[1]{}

\newcounter{Comment}[Comment]
\DeclareMathSymbol{\Delta}{\mathalpha}{operators}{1}
\DeclareMathSymbol{\Theta}{\mathalpha}{operators}{2}
\DeclareMathSymbol{\Pi}{\mathalpha}{operators}{5}
\DeclareMathSymbol{\Sigma}{\mathalpha}{operators}{6}

\newboolean{extended}      
\setboolean{extended}{false} %
\newcommand{\iflongpaper}[1]{\ifthenelse{\boolean{extended}}{#1}{}}
\newcommand{\ifregularpaper}[1]{\ifthenelse{\boolean{extended}}{}{#1}}

\newboolean{addthanks}      
\setboolean{addthanks}{false} %
\newcommand{\ifaddack}[1]{\ifthenelse{\boolean{addthanks}}{#1}{}}

{\noindent \textit{Proof sketch#1.}}{\mbox{}\nobreak\hfill\hspace{6pt}$\Box$}

\newcommand{\mxsat}{MaxSAT\xspace}
\newcommand{\wmxsat}{WMaxSAT\xspace}
\newcommand{\hmxsat}{HornMaxSAT\xspace}
\newcommand{\hwmxsat}{HornWMaxSAT\xspace}
\newcommand{\mxclq}{MaxClique\xspace}
\newcommand{\mxis}{MaxIS\xspace}
\newcommand{\mnvc}{MinVC\xspace}

\newcommand{\mnhs}{MinHS\xspace}
\newcommand{\mnsc}{MinSC\xspace}
\newcommand{\mnds}{MinDS\xspace}
\newcommand{\mxsp}{MaxSP\xspace}

\newcommand{\knap}{Knapsack\xspace}
\newcommand{\php}{PHP\xspace}

\newcommand{\mxhs}{MaxHS\xspace}
\newcommand{\hmxhs}{HMaxHS\xspace}

\newcommand{\elplus}{$\fml{EL}^{+}$\xspace}

\newcommand{\reduce}{\ensuremath \le_P}
\newcommand{\hencode}{\textsf{HEnc}}

\newcommand{\atmost}{\textsf{AtMost}}
\newcommand{\atleast}{\textsf{AtLeast}}

\graphicspath{{plots/}}
\DeclareGraphicsExtensions{.pdf}

\def\jpms{Joao Marques-Silva}
\def\alexi{Alexey Ignatiev}
\def\ajrm{Antonio Morgado}

\title{Horn Maximum Satisfiability:\\ Reductions, Algorithms \& Applications}
\titlerunning{Horn Maximum Satisfiability}

\author{
  {\jpms}\inst{1}
  \and
  {\alexi}\inst{1}
  \and
  {\ajrm}\inst{1}
}

\authorrunning{Marques-Silva, Ignatiev \& Morgado}

\institute{%
  LASIGE, Faculty of Science, University of Lisbon, Portugal\\
  \email{\{jpms,aignatiev,ajmorgado\}@ciencias.ulisboa.pt}
}

\AtBeginDocument{%

}

\begin {document}
\maketitle
\setcounter{footnote}{0}

%
%
%
%
\begin{abstract}
Recent years have witness remarkable performance improvements in
maximum satisfiability (\mxsat) solvers.
In practice, \mxsat algorithms often target the most generic \mxsat
formulation, whereas dedicated solvers, which address specific
subclasses of \mxsat, have not been investigated.
This paper shows that a wide range of optimization and decision
problems are either naturally formulated as \mxsat over Horn formulas,
or permit simple encodings using Horn \mxsat.
Furthermore, the paper also shows how linear time decision procedures
for Horn formulas can be used for developing novel algorithms for the
Horn \mxsat problem.
\end{abstract}
%

%
%
%
%

\section{Introduction} \label{sec:intro}

Recent years have seen very significant improvements in \mxsat solving 
technology~\cite{bacchus-cp11,mhlpms-cj13,ansotegui-aij13,jarvisalo-sat16}. 
Currently, the most effective \mxsat algorithms propose
different ways for iteratively finding and blocking unsatisfiable
cores (or subformulas).
However, and despite the promising results of \mxsat in practical
settings, past work has not investigated dedicated approaches for
solving subclasses of the \mxsat problem, with one concrete example
being the \mxsat problem over Horn formulas,
i.e.\ \hmxsat~\footnote{In contrast, for predicate logic and many of
  its specializations, Horn clauses are used ubiquitously. This
  includes logic programming, among many others applications.}.
%
The \hmxsat optimization problem is well-known to be
NP-hard~\cite{jaumard-ipl87}. In contrast to \hmxsat, the decision
problem for Horn formulas is well-known to be in P, with linear time
algorithms proposed in the 80s~\cite{gallier-jlp84,minoux-ipl88}.
This paper investigates practical uses of \mxsat subject to Horn
formulas, and shows that 
a vast number of decision and optimization problems are naturally
formulated as \hmxsat. More importantly, as this paper also shows, a
vast number of other decision and optimization problems admit simple
\hmxsat encodings.
%
One should observe that \hmxsat is NP-hard and so, by definition, any
decision problem in NP admits a polynomial time reduction to \hmxsat.
However, for many problems in NP, such reductions are not known, and
may result in large (even if polynomial) encodings.

With the purpose of exploiting the observation that many optimization
and decision problems have natural (and simple) reductions to \hmxsat,
this paper also proposes a novel algorithm for \hmxsat. The new
algorithm mimics
recent Implicit Hitting Set algorithms\footnote{Throughout the paper, these are referred to as \mxhs-family of \mxsat algorithms.} proposed for
\mxsat \cite{bacchus-cp11,jarvisalo-sat16}, thus exploiting the
fact that Horn formulas can be decided in polynomial (linear)
time~\cite{minoux-ipl88}, and for which minimal unsatisfiable cores
(or MUSes) can be computed in polynomial time~\cite{msimp-jelia16}.

The paper is organized as follows.~\autoref{sec:prelim} introduces the
definitions and notation used in the remainder of the
paper.~\autoref{sec:probs} shows that a large number of well-known
optimization, but also decision, problems already have simple \hmxsat
formulations which, to the best of our knowledge, have not been
exploited before.~\autoref{sec:algs} proposes a variant of recent
general-purpose \mxsat algorithms, that is dedicated to the \hmxsat
problem. This section also shows that the new algorithm can elicit
automatic abstraction mechanisms when solving large scale optimization
problems. The potential of the work proposed in this paper is
assessed in~\autoref{sec:res}, and~\autoref{sec:conc} concludes the
paper.


%
%
%

\section{Preliminaries} \label{sec:prelim}

The paper assumes definitions and notation standard in propositional
satisfiability (SAT) and \mxsat~\cite{sat-handbook09}.
Propositional variables are taken from a set $X=\{x_1,x_2,\ldots\}$.
A Conjunctive Normal Form (CNF) formula is defined as a conjunction of
disjunctions of literals, where a literal is a variable or its
complement. CNF formulas can also be viewed as sets of sets of 
literals, and are represented with calligraphic letters, $\fml{A}$,
$\fml{F}$, $\fml{H}$, etc.
Given a formula $\fml{F}$, the set of variables is
$\vars(\fml{F})\subseteq X$.
A clause is a \emph{goal clause} if all of its literals are negative.
A clause is a \emph{definite clause} if it has exactly one positive
literal and all the other literals are negative; the number of
negative literals may be 0. A clause is Horn if it is either a goal or
a definite clause.
A truth assignment $\nu$ is a map from variables to $\{0,1\}$. Given a 
truth assignment, a clause is satisfied if at least one of its
literals is assigned value 1; otherwise it is falsified. A formula is
satisfied if all of its clauses are satisfied; otherwise it is
falsified.
If there exists no assignment that satisfies a CNF formula $\fml{F}$,
then $\fml{F}$ is referred to as \emph{unsatisfiable}.
(Boolean) Satisfiability (SAT) is the decision problem for
propositional formulas, i.e.\ to decide whether a given propositional
formula is satisfiable.
Since the paper only considers propositional formulas in CNF,
throughout the paper SAT refers to the decision problem for
propositional formulas in CNF.
Modern SAT solvers instantiate the Conflict-Driven Clause Learning
paradigm~\cite{sat-handbook09}.
For unsatisfiable (or inconsistent) formulas, MUSes (minimal
unsatisfiable subsets) represent subset-minimal subformulas that are
unsatisfiable (or inconsistent), and MCSes (minimal correction
subsets) represent subset-minimal subformulas such that the complement
is satisfiable~\cite{sat-handbook09}.

To simplify modeling with propositional logic, one often represents
more expressive constraints. Concrete examples are cardinality
constraints and pseudo-Boolean constraints~\cite{sat-handbook09}.
A cardinality constraint of the form $\sum x_i\le k$ is referred to as
an $\atmost{}k$ constraint, whereas a cardinality constraint of the
form $\sum x_i\ge k$ is referred to as an $\atleast{}k$ constraint.
Propositional encodings of cardinality and pseudo-Boolean constraints
is an area of active
research~\cite{warners-ipl98,bailleux-cp03,sinz-cp05,een-jsat06,sat-handbook09,nieuwenhuis-sat09,codish-lpar10,roussel-sat09,nieuwenhuis-cj11,nieuwenhuis-sat11,koshimura-ictai13a}.

The (plain) \mxsat problem is to find a truth assignment that
maximizes the number of satisfied clauses. For the plain \mxsat
problem, all clauses are \emph{soft}, meaning that these may not be
satisfied. Variants of the \mxsat can consider the existence of
\emph{hard} clauses, meaning that these must be satisfied, and also
assign weights to the soft clauses, denoting the \emph{cost} of
falsifying the clause; this is referred as the weighted \mxsat
problem, \wmxsat. When addressing \mxsat problems with weights, hard
clauses are assigned a large weight $\top$.
The \hmxsat problem corresponds to the \mxsat problem when all clauses
are Horn. If clauses have weights, then \hwmxsat denotes the Horn
\mxsat problem when the soft clauses have weights.

Throughout the paper, standard graph and set notations will be used.
An undirected graph $G=(V,E)$ is defined by a set $V$ of vertices
and a set $E\subseteq\{\{u,v\}\,|\,u,v\in V, u\not=v\}$. The notation
$(u,v)$ is used in this paper to represent the edges $\{u,v\}$ of $E$,
where the order of the vertices is irrelevant. Given $G=(V,E)$, the
\emph{complement graph} $G=(V,E^C)$ is the graph with the edges in
$\{\{u,v\}\,|\,u,v\in V, u\not=v\}$ that are not in $E$.
Moreover, it is assumed some familiarity with optimization problems
defined on graphs, including minimum vertex cover, maximum independent
set, maximum clique, among others.
Finally, the notation $\reduce$ is used to represent polynomial time
reducibility between problems~\cite[Section~34.3]{cormen-bk09}.


%
%
%

\section{Basic Reductions} \label{sec:probs}

This section shows that a number of well-known problems can be reduced
in polynomial time to the \hmxsat problem. Some of the reductions are
well-known; we simply highlight that the resulting propositional
formulas are Horn.

\subsection{Optimization Problems on Graphs}

%
\begin{ndefinition}[Minimum Vertex Cover, \mnvc]
  Given an undirected graph $G=(V,E)$, a vertex cover $T\subseteq V$
  is such that for each $(u,v)\in E$, $\{u,v\}\cap T\not=\emptyset$.
  A minimum (or cardinality minimal) vertex cover $T\subseteq V$ is a
  vertex cover of minimum size\footnote{This corresponds to requiring
    $T\subseteq V$ to be such that
    $\forall_{U\subseteq V}|U|<|T|\to\exists_{(u,v)\in E},\{u,v\}\cap U=\emptyset$.
    Throughout the paper, we will skip the mathematical representation
    of minimum (but also maximum) size sets.}.
\end{ndefinition}
\begin{nreduction}[$\text{\mnvc}\reduce\text{\hmxsat}$]
  For $u\in V$, let $x_u=1$ iff $u$ is \emph{not} included in a vertex
  cover. For any $(u,v)\in E$, 
  add a hard clause $(\neg x_u\lor \neg x_v)$. For each $u\in V$, add
  a soft clause $(x_u)$. (Any non-excluded vertex $u\in V$ 
  (i.e.\ $x_u=0$) is in the vertex cover.)
\end{nreduction}

\begin{nremark}
  The proposed reduction differs substantially from the one originally
  used for proving \hmxsat to be NP-hard~\cite{jaumard-ipl87}, but our
  working assumptions are also distinct, in that we consider hard and
  soft clauses. 
\end{nremark}

%
\begin{ndefinition}[Maximum Independent Set, \mxis]
  Given an undirected graph $G=(V,E)$, an independent set
  $I\subseteq V$ is such that for each $(u,v)\in E$ either
  $u\not\in I$ or $v\not\in I$.
  A maximum independent set is an independent set of maximum size.
\end{ndefinition}
\begin{nreduction}[$\text{\mxis}\reduce\text{\hmxsat}$]
  One can simply use the previous encoding, by noting the relationship
  between vertex covers and independent sets. For any $(u,v)\in E$,
  add a hard clause $(\neg x_u\lor\neg x_v)$. For each $u\in V$, add
  a soft clause $(x_u)$.
\end{nreduction}

%
\begin{ndefinition}[Maximum Clique, \mxclq] \label{def:mxclq}
  Given an undirected graph $G=(V,E)$, a clique (or complete subgraph)
  $C\subseteq V$ is such that for every pair $\{u,v\}\subseteq C$,
  $(u,v)\in E$.
  A maximum clique is a clique of maximum size.
\end{ndefinition}
\begin{nreduction}[$\text{\mxclq}\reduce\text{\hmxsat}$]
  A \mxsat encoding for \mxclq is the following. For any $(u,v)\in
  E^{C}$, add a hard clause $(\neg x_u\lor\neg x_v)$. For each $u\in
  V$, add a soft clause $(x_u)$.
\end{nreduction}

%
\begin{ndefinition}[Minimum Dominating Set, \mnds] \label{def:mnds}
  Let $G=(V,E)$ be an undirected graph. $D\subseteq V$ is a dominating
  set if any $v\in V\setminus D$ is adjacent to at least one vertex in
  $D$.
  A minimum dominating set is a dominating set of minimum size.
\end{ndefinition}
\begin{nreduction}[$\text{\mnds}\reduce\text{\hmxsat}$]
  Let $x_u=1$ iff $u\in V$ is excluded from a dominating set $D$.
  For each vertex $u\in V$ add a hard Horn clause
  $(\neg x_u\lor_{(u,v)\in E}\neg x_v)$.
  The soft clauses are $(x_u)$, for $u\in V$.
\end{nreduction}


%

\subsection{Optimization Problems on Sets}

%
\begin{ndefinition}[Minimum Hitting Set, \mnhs] \label{def:mnhs}
  Let $\fml{C}$ be a collection of sets of some set $S$. A hitting set
  $H\subseteq S$ is such that for any $D\in\fml{C}$,
  $H\cap D\not=\emptyset$.
  A minimum hitting set is a hitting set of minimum size.
\end{ndefinition}
\begin{nreduction}[$\text{\mnhs}\reduce\text{\hmxsat}$]
  For each $a\in S$ let $x_a=1$ iff $a$ is excluded from $H$.
  For each $D\in\fml{C}$, create a hard Horn clause
  $(\lor_{a\in D}\neg x_a)$.
  The soft clauses are $(x_a)$, for $a\in S$.
\end{nreduction}

\begin{nremark} The minimum set cover (\mnsc) is well-known to be
  equivalent to the minimum hitting set problem. Thus, the same
  reduction to \hmxsat can be applied.
\end{nremark}

\begin{ndefinition}[Maximum Set Packing, \mxsp] \label{def:mxsp}
  Let $\fml{T}=\{T_1,\ldots,T_k\}$ be a family of sets.
  $\fml{R}\subseteq\fml{T}$ is a set packing if
  $\forall_{T_i,T_j\in\fml{R}}T_i\cap T_j=\emptyset$. A maximum set
  packing is a set packing of maxim size.
\end{ndefinition}
\begin{nreduction}[$\text{\mxsp}\reduce\text{\hmxsat}$]
  Let $x_{i}=1$ iff $T_i$ is included in the set packing.
  For each pair $T_i,T_j$, such that $T_i\cap T_j\not=\emptyset$,
  create a hard Horn clause $(\neg x_i\lor \neg x_j)$. The soft
  clauses are $(x_i)$, for $T_i\in\fml{T}$.
\end{nreduction}

\begin{nremark}
  It is well-known that the maximum set packing problem can be reduced
  to the maximum clique problem. The reduction above exploits this
  result.
\end{nremark}

It also immediate to conclude that the weighted version of any of the
optimization problems described in this and the previous sections can  
be reduced to \hwmxsat.

\subsection{Handling Linear Constraints} \label{ssec:prob3}

This section argues that the propositional encodings of a number of
linear constraints are Horn. In turn, this enables solving a number of
optimization problems with \hmxsat.

The first observation is that \emph{any} of the most widely used CNF
encodings of AtMost$k$ constraints are composed \emph{exclusively} of
Horn clauses\footnote{To our best knowledge, this property of
  propositional encodings has not been investigated before.}:
\begin{nproposition}[CNF Encodings of AtMost$k$ constraints]
  \label{prop:cardhorn}
  The following CNF encodings of AtMost$k$ constraints are composed
  solely of Horn clauses:
  pairwise and bitwise encodings~\cite[Chapter~2]{sat-handbook09},
  totalizers~\cite{bailleux-cp03}, sequential
  counters~\cite{sinz-cp05}, sorting networks~\cite{een-jsat06},
  cardinality networks~\cite{nieuwenhuis-sat09,nieuwenhuis-cj11},
  pairwise cardinality networks~\cite{codish-lpar10}, and
  modulo totalizers~\cite{koshimura-ictai13a}.
\end{nproposition}
\begin{proof}
  Immediate by inspection of each
  encoding~\cite{sat-handbook09,bailleux-cp03,sinz-cp05,een-jsat06,nieuwenhuis-sat09,codish-lpar10,nieuwenhuis-cj11,koshimura-ictai13a}.
\end{proof}

For the case of the more general pseudo-Boolean constraints,
$\sum a_i x_i\le b$, with $a_i, b$ non-negative, there also exist Horn
encodings:
\begin{nproposition}[CNF Encodings of Pseudo-Boolean Constraints]\label{prop:pbhorn}
  The Local Polynomial Watchdog~\cite{roussel-sat09} encoding for PB
  constraints is 
  composed solely of Horn clauses.
\end{nproposition}
\begin{proof}
  Immediate by inspection of the encoding in~\cite{roussel-sat09}.
\end{proof}

\jnoteF{Cardinality constraints, including AtMost1 constraints, have
  Horn
  encodings~\cite{bailleux-cp03,sinz-cp05,een-jsat06,nieuwenhuis-sat09,codish-lpar10,nieuwenhuis-cj11,koshimura-ictai13a}. More
  importantly, pseudo-Boolean constraints also have Horn
  encodings~\cite{roussel-sat09}. All these encodings
  guarantee arc-consistency.}

These observations have immediate impact on the range of problems that
can be solved with \hmxsat and \hwmxsat. One concrete example is the
Knapsack problem~\cite{cormen-bk09}.
%
%
\begin{ndefinition}[Knapsack problem] \label{def:}
  Let $\{1,\ldots,n\}$ denote a set of $n$ objects, each with value
  $v_i$ and weight $w_i$, $1\le i\le n$, and a maximum weight value
  $W$. The knapsack problem is to pick a subset of objects of maximum
  value that is consistent with the weight constraint. By letting
  $x_i=1$ iff object $i$ is picked, we get the well-known 0-1 ILP
  formulation $\text{max}\sum_{i}v_ix_i;\:\text{s.t.}\sum_{i}w_ix_i\le
  W$.
\end{ndefinition}
\begin{nreduction}[$\text{\knap}\reduce\text{\hmxsat}$]
  From~\autoref{prop:pbhorn}, there exist Horn encodings for
  Pseudo-Boolean constraints.
  Thus, the hard constraint   $\sum_{i}w_ix_i\le W$ can be encoded
  with Horn clauses.
  The soft clauses are $(x_i)$ for each object $i$, each with cost
  $v_i$.
  Both the soft and the hard clauses in the reduction are Horn.
\end{nreduction}


%
%
%

\section{HornMaxSAT Algorithm with Hitting Sets} \label{sec:algs}

This section develops a \mxhs-like~\cite{bacchus-cp11,jarvisalo-sat16}
algorithm for \hmxsat. In addition, the section shows that this
\mxhs-like algorithm elicits the possibility of solving large scale
problems with abstraction.

\subsection{A \mxhs-Like \hmxsat Algorithm}

With the goal of exploiting the special structure of \hmxsat, a
\mxhs-like algorithm is
envisioned~\cite{bacchus-cp11,jarvisalo-sat16}.
\begin{algorithm}[t]
%
%
%
%
%
%
%

\DontPrintSemicolon
\SetAlgoNoLine
\LinesNumbered
\SetFillComment
\SetKw{KwNot}{not\xspace}
\SetKw{KwAnd}{and\xspace}
\SetKw{KwOr}{or\xspace}
\SetKw{KwBreak}{break\xspace}
\SetKwData{false}{{\small false}}
\SetKwData{true}{{\small true}}
\SetKwData{st}{{\slshape st}}
\SetKwData{cores}{$\mathcal{C}$}
\SetKwFunction{LTUR}{LTUR}
\SetKwFunction{SAT}{SAT}
\SetKwFunction{CNF}{CNF}
\SetKwFunction{minhs}{MinimumHS}
\SetKwFunction{getmus}{ComputeMUS}
\SetKwFunction{relaxcls}{RelaxCls}
\SetKwFunction{softcls}{SoftCls}
\SetKwFunction{falsecls}{PickFalseCls}
\SetKwFunction{relaxvars}{GetRelaxationVars}
\SetKwFunction{ip}{IP}
\SetKwFunction{cost}{Cost}
\SetKwBlock{Let}{let}{end}
\SetKwBlock{FBlock}{}{end}

\KwIn{$\fml{F}=\langle\fml{A}, \fml{H}\rangle$, \hmxsat formula}
\KwOut{$(\mu, \cost(\mu))$, MaxSAT assignment and cost}
\Begin{
  \SetAlgoVlined
  $\fml{K}\gets \emptyset$ \;
  \While{\true}{
    $\fml{S} \gets \minhs(\fml{K})$ \;
    $(\st, \mu, \fml{U})\gets\LTUR(\fml{H}\cup(\fml{A}\setminus\fml{S}))$ \;
    \tcp{If $\st$, then $\mu$ is a satisfying assignment}
    \tcp{Otherwise, $\fml{U}$ is a core/MUS}
    \lIf{$\st$}{\Return $(\mu, \cost(\mu))$}
    $\fml{K}\gets\fml{K} \cup \{\fml{U}\}$\;
  }
}
\BlankLine
%

  \caption{\hmxhs, a \mxhs-like~\cite{bacchus-cp11} \hmxsat algorithm}
  \label{alg:hmaxhs} 
\end{algorithm}
\autoref{alg:hmaxhs} summarizes the proposed approach. The key
observation is that each call to LTUR~\cite{minoux-ipl88} runs in
linear time. (Unit propagation as implemented in modern SAT solvers,
will also run in polynomial time, but it will be less efficient in
practice.) The original motivation for MaxHS is that finding a
minimum hitting set of $\fml{S}$ is expected to be much easier than
solving the \mxsat problem. This is also the motivation for \hmxhs.
As observed in recent work~\cite{amms-sat15,msimp-jelia16}, MUSes
(minimal unsatisfiable subsets) can be computed in polynomial time in
the case of Horn formulas. MUS extraction, but also MCS (minimal
correction subset) extraction~\cite{msimp-jelia16}, are based on the
original LTUR algorithm~\cite{minoux-ipl88}.
It should be noted that some implementations of \mxhs use an ILP
(Integer Linear Programming) package (e.g.\ CPLEX or
SCIP)~\cite{bacchus-cp11,jarvisalo-sat16}\footnote{SCIP and CPLEX and
  available, respectively, from \url{http://scip.zib.de/} and
  \url{https://www-01.ibm.com/software/commerce/optimization/cplex-optimizer/}.},
whereas others exploit SAT solvers for computing minimum hitting
sets~\cite{iplms-cp15,imms-ecai16}.

\subsection{Automatic Abstraction-Based Problem Solving}
\label{sec:abstract}

For some of the problems described in~\autoref{sec:probs} a possible
criticism of~\autoref{alg:hmaxhs} is that it will iteratively find
sets $\fml{U}$ consisting of a single clause, and it will essentially
add to $\fml{K}$ all the clauses in $\fml{H}$. Although this is in
fact a possibility for some problems (but not all, as investigated
in~\autoref{sec:apps}), this section shows that even for these
problems,~\autoref{alg:hmaxhs} can provide an effective problem
solving approach.

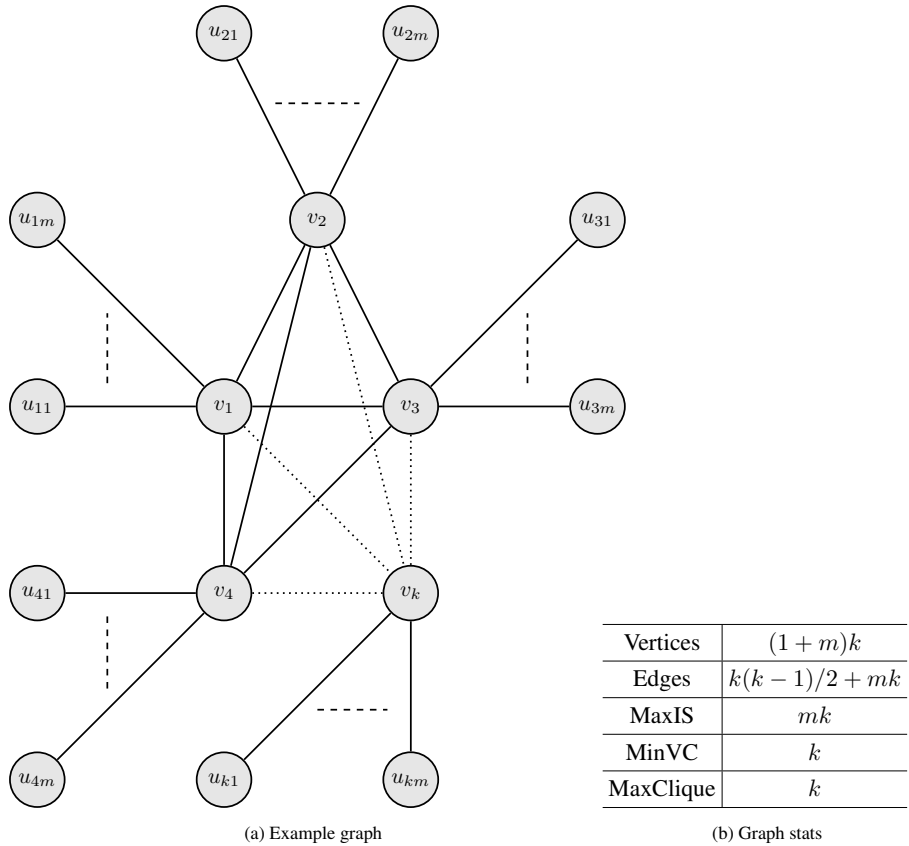
\begin{figure}[t]
 \begin{subfigure}[b]{0.675\textwidth}
   \scalebox{0.8275}{
%
%
%
%
\tikzset{
    every node/.style={
        circle,
        draw,
        fill          = black!50,
        inner sep     = 0pt,
        minimum width =4 pt
    }   
}  

\begin{tikzpicture}[thick,scale=1.5,-,
    every node/.style={circle,
      draw,
      fill          = black!10,
      inner sep     = 2.5pt,
      minimum width =2.5em}]
  \path[draw] 

     node (v1) at (2,4) {$v_1$}
     node (v2) at (3,6) {$v_2$}
     node (v3) at (4,4) {$v_3$}
     node (v4) at (2,2) {$v_4$}
     node (vk) at (4,2) {$v_k$}
     %
     node (uk1) at (2,0) {$u_{k1}$}
     node (ukm) at (4,0) {$u_{km}$}
     node (u41) at (0,2) {$u_{41}$}
     node (u4m) at (0,0) {$u_{4m}$}
     node (u31) at (6,6) {$u_{31}$}
     node (u3m) at (6,4) {$u_{3m}$}
     node (u21) at (2,8) {$u_{21}$}
     node (u2m) at (4,8) {$u_{2m}$}
     node (u11) at (0,4) {$u_{11}$}
     node (u1m) at (0,6) {$u_{1m}$}
     ;

     \draw (v1) -- (v1) ; 
     \draw (v1) -- (v2) ;
     \draw (v1) -- (v3) ;
     \draw (v1) -- (v4) ;
     \draw (v1) -- (u11) ;
     \draw (v1) -- (u1m) ;
     \draw (v2) -- (v3) ;
     \draw (v2) -- (v4) ;
     \draw (v2) -- (u21) ;
     \draw (v2) -- (u2m) ;
     \draw (v3) -- (v4) ;
     \draw (v3) -- (u31) ;
     \draw (v3) -- (u3m) ;
     %
     \draw (v4) -- (u41) ;
     \draw (v4) -- (u4m) ;
     \draw (vk) -- (uk1) ;
     \draw (vk) -- (ukm) ;

\begin{scope}   [dashed]  
     \draw (0.75,1.75) -- (0.75,0.95) ;
     \draw (0.75,4.25) -- (0.75,5.0) ;
     \draw (2.55,7.25) -- (3.45,7.25) ;
     \draw (3.0,0.75) -- (3.75,0.75) ;
     \draw (5.25,4.25) -- (5.25,5.0) ;
\end{scope}

\begin{scope}   [dotted]  
     \draw (v1) -- (vk)  ; 
     \draw (v2) -- (vk)  ;
     \draw (v3) -- (vk)  ;
     \draw (v4) -- (vk)  ;
\end{scope}

\end{tikzpicture}
%
}
   \caption{Example graph} \label{fig:ex01-graph}
 \end{subfigure}
 \begin{subfigure}[b]{0.3\textwidth}
   \scalebox{0.9}{
     
     \renewcommand{\tabcolsep}{0.25em}
     \hspace*{-0.75cm}
     \begin{tabular}{c|c} \hline
       Vertices & $(1+m)k$ \\ \hline 
       Edges & $k(k-1)/2+mk$ \\ \hline
       \mxis & $mk$ \\ \hline
       \mnvc & $k$ \\ \hline
       \mxclq & $k$ \\ \hline
     \end{tabular}
   }
   \caption{Graph stats} \label{fig:ex01-stats}
 \end{subfigure}
 \caption{Example graph for computing \mxis \& \mnvc} \label{fig:graph01}
\end{figure}

Consider the example graph in~\autoref{fig:graph01}, where the goal is
to compute a maximum independent set (or alternatively a minimum
vertex cover).
From the figure, we can conclude that the number of vertices is
$(1+m)k$, the number of edges is $(k(k-1)/2+km)$, the size of the
maximum independent set is $km$ and the size of the minimum vertex
cover is $k$.
From the inspection of the reduction from \mxis (or \mnvc) to \hmxsat,
and the operation of~\autoref{alg:hmaxhs}, one might anticipate
that~\autoref{alg:hmaxhs} would iteratively declare each hard clause
as an unsatisfiable core, and replicate the clause in the list
$\fml{K}$ of sets to hit, thus requiring a number of iterations no
smaller than the number of edges. (More importantly, for a \mxhs-like
algorithm, the number of iterations is worst-case
exponential~\cite{bacchus-cp11}.)
However, and as shown below, the operation of the \hmxhs actually
\emph{ensures} this is \emph{not} the case.
%
%

Without loss of generality, consider any of the vertices in the
clique, i.e.\ $v_1,\ldots, v_k$, say $v_i$. For this vertex, no more
than $k(k-1)/2+2k$ edges will be replicated, i.e.\ added to $\fml{K}$.
Observe that, as soon as two edges $(v_i, u_{ij_1})$ and
$(v_i,u_{ij_2})$ are replicated, a minimum hitting set will
necessarily pick $v_i$. As a result, after at most $k(k-1)/2+2k$
iterations, the algorithm will terminate with the answer $mk$.
Essentially, the algorithm is capable of \emph{abstracting} away
$(m-2)k$ clauses when computing the maximum independent set. Observe
that $m$ can be made arbitrarily large.
Abstraction is a well-known topic in AI, with important
applications~\cite{walsh-aij92}. The example in this section suggests
that \hmxsat and the \hmxhs algorithm can effectively enable automatic
abstraction for solving large scale (graph) optimization
problems. This remark is further investigated in~\autoref{sec:res}.

It should be noted that the result above highlights what seems to be a
fundamental property of the original \mxhs
algorithm~\cite{bacchus-cp11}.
Although in the worst case, the algorithm can require an exponential
number of steps to find the required set of clauses to remove to
achieve consistency, the result above illustrates how the \mxhs can be
effective at discarding irrelevant clauses, and focusing on the key
parts of the formula, thus being able to compute solutions in a number
of iterations not much larger than the minimum number of falsified
clauses in the \mxhs solution. Practical results from recent \mxsat
Evaluations\footnote{\url{http://www.maxsat.udl.cat/}.} confirm the
practical effectiveness of \mxhs-like algorithms.


%
%
%

\section{HornMaxSAT in Practice} \label{sec:apps}

Besides the reference optimization problems analyzed
in~\autoref{sec:probs}, a number of practical applications can also be
shown to correspond to solving \hmxsat or can be reduced to \hmxsat.
This section investigates some of these problems, but also proposes
generic encodings from either SAT and CSP into \hmxsat.

\subsection{Sample Problems}
Different optimization problems in practical settings are encoded as
\hmxsat.

The winner determination problem (WDP) finds important applications in
combinatorial auctions. An immediate observation is that the encoding
proposed in~\cite{larrosa-jsat08} corresponds to \hmxsat.
The problem of coalition structure generation (CSG) also finds
important applications in multi-agent systems. An immediate
observation is that some of the encodings proposed
in~\cite{koshimura-ictai12} correspond to \hmxsat.
\hmxsat also finds application in the area of axiom pinpointing for
\elplus description logic, but also for other lightweight description
logics.
For the concrete case of \elplus, the problem encoding is well-known
to be Horn~\cite{sebastiani-cade09}, with the soft clauses being unit
positive. The use of LTUR-like algorithms has been investigated
in~\cite{amms-sat15}. 
%
%
%

\jnoteF{
  Goal models \& Coalition struture generation.\\
  Detail model for winner determination problem.}

As shown in the sections below, it is actually simple to map different
decision (and optimization\footnote{
  In the case of optimization problems, it is simple to apply the same
  technique in the setting of Boolean Lexicographic Optimization
  (BLO)~\cite{msagl-amai11}. Due to lack of space, details are
  mitted.}) problems into \hmxsat.

\subsection{Reducing SAT to \hmxsat} \label{ssec:sat2horn}

Let $\fml{F}$ be a CNF formula, with $N$ variables $\{x_1\ldots,x_N\}$
and $M$ clauses $\{c_1,\ldots,c_M\}$.
Given $\fml{F}$, the reduction creates a Horn \mxsat problem with hard
clauses $\fml{H}$ and soft clauses $\fml{S}$,
$\langle\fml{H},\fml{S}\rangle=\hencode(\fml{F})$.
For each variable $x_i\in X$, create new variables $p_i$ and $n_i$,
where $p_i=1$ iff $x_i=1$, and $n_i=1$ iff $x_i=0$. Thus, we need a hard
clause $(\neg p_i\lor\neg n_i)$, to ensure that we do not
simultaneously assign $x_i=1$ and $x_i=0$. (Observe that the added
clause is Horn.)
For each clause $c_j$ we require  $c_j$ to be satisfied, by requiring 
that one of its literals \emph{not} to be falsified. For each literal
$x_i$ use $\neg n_i$ and for each literal $\neg x_i$ use $\neg p_i$.
Thus, $c_j$ is encoded with a new (hard) clause $c'_j$  with the same
number of literals as $c_i$, but with only negative literals on the
$p_i$ and $n_i$ variables, and so the resulting clause is also Horn. 
The set of soft clauses $\fml{S}$ is given by $(p_i)$ and $(n_i)$ for
each of the original variables $x_i$.
If the resulting Horn formula has a \hmxsat solution with at least
$N$ variables assigned value 1, then the original formula is
satisfiable; otherwise the original formula is unsatisfiable. (Observe
that, by construction, the \hmxsat solution cannot assign value 1 to
more than $N$ variables.)
Clearly, the encoding outlined in this section can be the subject of
different improvements, e.g.\ not all clauses need to be goal
clauses. 

The transformation proposed can be related with the well-known
dual-rail encoding, used in different
settings~\cite{BryantBBCS87,mfmso-ictai97,RoordaC05,jmsss-jelia14,pimms-ijcai15}. To
our best knowledge, the use of a dual-rail encoding for deriving a
pure Horn formula has not been proposed in earlier work.

\subsection{Reducing CSP to \hmxsat} \label{ssec:csp2horn}

This section investigates reductions of Constraint Satisfaction
Problems (CSP) into \hmxsat. Standard definitions are
assumed~\cite{walsh-bk06}.
A CSP is a triple $\langle X, D, C\rangle$, where $X=\langle
x_1,\ldots,x_N\rangle$ is an $n$-tuple of variables, $D$ is a
corresponding $N$-tuple of domains $D=\langle D_1,\ldots,D_N\rangle$,
such that $x_i\in D_i$, and $C$ is a $t$ tuple of constraints $C=\langle
C_1,\ldots,C_t\rangle$. $C_j$ is a pair $\langle R_{S_j}, S_j\rangle$,
where $R_{S_j}$ is a relation on the variables in $S_j$, and
represents a subset of the cartesian product of the domains of the
variables in $S_j$.

One approach to encode CSPs as \hmxsat is to translate the CSP to SAT
(e.g.\cite{walsh-cp00}), and then apply the Horn encoder outlined
in~\autoref{ssec:sat2horn}. There are however, alternative approaches,
one of which we now detail.
We show how to adapt the well-known direct encoding of CSP into
SAT~\cite{walsh-cp00}. The set of variables is $x_{iv}$, such that
$x_{iv}=1$ iff $x_i$ is assigned value $v\in D_i$. Moreover, we
consider the \emph{disallowed} combinations of values of each
constraint $C_j$. For example, if the combination of values
$x_{i_1}=v_{i_1}\land x_{i_2}=v_{i_2}\land\cdots\land x_{i_q}=v_{i_q}$
is disallowed, i.e.\ no tuple of the relation $S_j$ associated with
$C_j$ contains these values, then add a (Horn) clause
$(\neg x_{i_1v_{i_1}}\lor\cdots\lor\neg x_{i_qv_{i_q}})$.
For each $x_i$, require that no more than one value can be used:
$\sum_{v\in D_i}x_{iv}\le 1$; this AtMost1 constraint can be encoded
with Horn clauses as shown in~\autoref{prop:cardhorn}.
Finally, the goal is to assign as many variables as possible, and so
add a soft clause $(x_{i,v})$ for each $x_i$ and each $v\in D_i$.
It is immediate that the CSP is satisfiable iff the \hmxsat
formulation has a solution with at least $N$ satisfied soft clauses
(and by construction it cannot assign value 1 to more than $N$
variables).

\jnoteF{Detail the CSP2Horn SAT encoding.}


\subsection{Reducing PHP to \hmxsat}

The previous sections show that the optimization and decision problems 
with simple reductions to \hmxsat are essentially endless,
as any decision problem that can be reduced to SAT or CSP can also
be reduced to \hmxsat.
However, it is also possible to develop specific reductions, that
exploit the original problem formulation. This section investigates
how to encode the representation of the pigeonhole principle (PHP) as
\hmxsat, for which propositional encodings are well-known and
extensively investigated~\cite{cook-jsl79}.

\begin{ndefinition}[Pigeonhole Principle, \php~\cite{cook-jsl79}]
  The pigeonhole principle states that if $m+1$ pigeons are
  distributed by $m$ holes, then at least one hole contains more than
  one pigeon. A more formal formulation is that  there exists no
  injective function mapping $\{1,2,...,m+1\}$ to $\{1,2,...,m\}$, for
  $m\ge1$.
\end{ndefinition}
Propositional formulations of $\text{\php}$ encode the negation of the
principle, and ask for an assignment such that the $m+1$ pigeons are
plansed into $m$ holes~\cite{cook-jsl79}. Given a propositional
encoding and the reduction proposed in~\autoref{ssec:sat2horn}, we can
encode PHP formulas into \hmxsat. We describe below an alternative
reduction.

\begin{nreduction}[$\text{\php}\reduce\text{\hmxsat}$]
  Let $x_{ij}=1$ iff pigeon $i$, with $1\le i\le m+1$, is placed in
  hole $j$, with $1\le j\le m$.
  For each hole $j$, $1\le j\le m$, at most 1 pigeon can be placed in
  hole $j$:
  \begin{equation} \label{eq:amo1}
    \begin{array}{lcl}
      \sum_{i=1}^{m+1}x_{ij}\le 1 & \quad\quad\quad & 1\le j\le m \\
    \end{array}
  \end{equation}
  which can be encoded with Horn clauses, by~\autoref{prop:cardhorn}.\\
  For each pigeon $i$, $1\le i\le m+1$, the pigeon is placed in at most 1
  hole:
  \begin{equation}  \label{eq:amo2}
    \begin{array}{lcl}
      \sum_{j=1}^{m}x_{ij}\le 1 & \quad\quad\quad & 1\le i \le m+1 \\
    \end{array}
  \end{equation}
  which can also be encoded with Horn clauses,
  by~\autoref{prop:cardhorn}.\\
  The soft clauses are $(x_{ij})$, with $1\le i\le m+1, 1\le j\le m$.
  The PHP problem is satisfiable iff the \hmxsat problem has a
  solution satisfying at least $m+1$ soft clauses, i.e.\ $m+1$ are
  placed. 
\end{nreduction}


%
%
%

\section{Experimental Results} \label{sec:res}



\begin{figure}[!t]
  \begin{center}
    \includegraphics[width=0.85\textwidth]{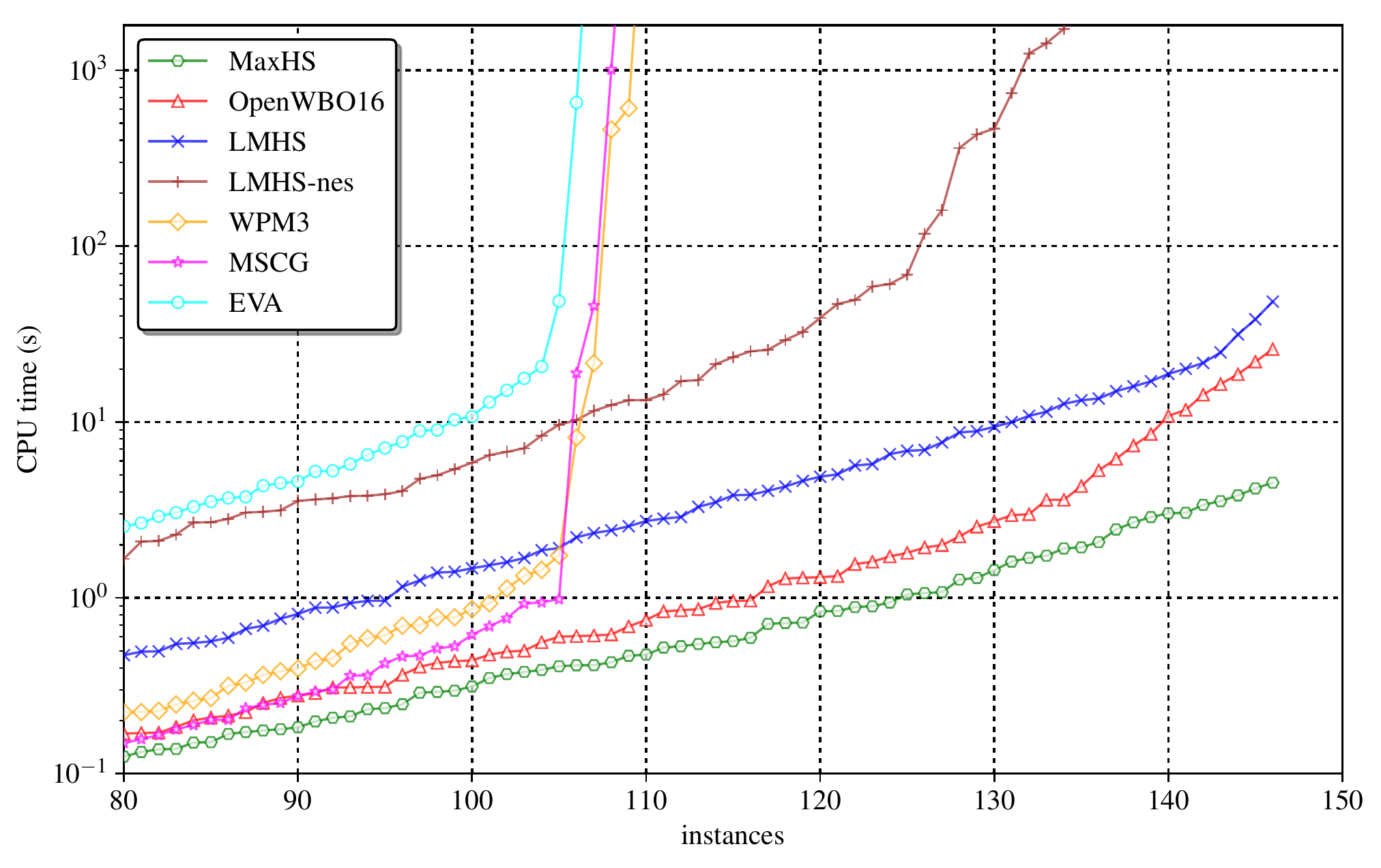}
    \caption{Cactus plot for selected solvers on \php and \mxis benchmarks.}
    \label{fig:cactus}
  \end{center}
\end{figure}

This section provides a preliminary investigation into exploiting
reductions to \hmxsat in practice.
All the experiments were run in Ubuntu Linux on an Intel Xeon~E5-2630
2.60GHz processor with 64GByte of memory. The time limit was set to
1800s and the memory limit to 10GByte for each process to run.
Two classes of problem instances were considered.
The first being a set of 46 \php instances that were generated ranging
the number of holes from 10 up to 100.
The second set of benchmarks corresponds to 100 instances generated
according to the example in \autoref{fig:graph01}, with $k$ ranging
from 10 to 100 and $m$ ranging from $k$ to $20k$. 
In the experiments six different \mxsat solvers were considered.
Some solvers are core-guide~\cite{mhlpms-cj13} (namely, OpenWBO16,
WPM3, MSCG and Eva), whereas others are based on implicit hitting sets
(namely, MaxHS and LMHS)~\cite{mhlpms-cj13}.
Additionally, a variant of LMHS was considered for which the option
"--no-equiv-seed" was set (LMHS-nes).
The results are summarized in the cactus plot shown
in~\autoref{fig:cactus}.
As can be observed, solvers based on implicit hitting sets (i.e.\ the
MaxHS family of \mxsat algorithms), but also OpenWBO16, perform very
well one the instances considered\footnote{Any implementation of the
  MaxHS-family of \mxsat algorithms, by using a CDCL SAT solver,
  implements a basic version of the algorithm proposed
  in~\autoref{sec:algs}. 
}.
The differences to the other solvers are solely due to the PHP
instances.
While propositional encodings of PHP are well-known to be extremely
hard for SAT solvers, the proposed \mxsat encoding scales well for
\mxhs-like algorithms, but also for the core-guided \mxsat solver
OpenWBO16. 

\begin{table}[t]
  \begin{center}
    \caption{Statistics on benchmarks generated according to the example in \autoref{fig:graph01}.} \label{tab:maxis-numbers}
%
%
%
%

\renewcommand{\tabcolsep}{0.325em}
{\scriptsize
\begin{tabular}{c|c|c|c|c|c|c|c|c|c|c|c|c|c|c|c|c|c|c}
\hline
\scriptsize $k$ & \multicolumn{2}{c|}{10}
    & \multicolumn{2}{c|}{20}
    & \multicolumn{2}{c|}{30}
    & \multicolumn{2}{c|}{40}
    & \multicolumn{2}{c|}{50}
    & \multicolumn{2}{c|}{60}
    & \multicolumn{2}{c|}{70}
    & \multicolumn{2}{c|}{80}
    & \multicolumn{2}{c}{90} \\
\hline
\scriptsize $m$ & \scriptsize 100 & \scriptsize 200
    & \scriptsize 200 & \scriptsize 400
    & \scriptsize 300 & \scriptsize 600
    & \scriptsize 400 & \scriptsize 800
    & \scriptsize 500 & \scriptsize 1000
    & \scriptsize 600 & \scriptsize 1200
    & \scriptsize 700 & \scriptsize 1400
    & \scriptsize 800 & \scriptsize 1600
    & \scriptsize 900 & \scriptsize 1800 \\
\hline
\scriptsize UB
   & \scriptsize 65 & \scriptsize 65
   & \scriptsize 230 & \scriptsize 230
   & \scriptsize 495 & \scriptsize 495
   & \scriptsize 860 & \scriptsize 860
   & \scriptsize 1325 & \scriptsize 1325
   & \scriptsize 1890 & \scriptsize 1890
   & \scriptsize 2555 & \scriptsize 2555
   & \scriptsize 3320 & \scriptsize 3320
   & \scriptsize 4185 & \scriptsize 4185 \\
\hline
\scriptsize \#DC
     & \scriptsize 9 & \scriptsize 7
     & \scriptsize 13 & \scriptsize 13
     & \scriptsize 27 & \scriptsize 26
     & \scriptsize 25 & \scriptsize 25
     & \scriptsize 50 & \scriptsize 50
     & \scriptsize 49 & \scriptsize 36
     & \scriptsize 70 & \scriptsize 70
     & \scriptsize 48 & \scriptsize 49
     & \scriptsize 63 & \scriptsize 63\\
\hline
\scriptsize \#I
    & \scriptsize 19 & \scriptsize 35
    & \scriptsize 71 & \scriptsize 132
    & \scriptsize 53 & \scriptsize 72
    & \scriptsize 211 & \scriptsize 356
    & \scriptsize 50 & \scriptsize 50
    & \scriptsize 225 & \scriptsize 693
    & \scriptsize 70 & \scriptsize 70
    & \scriptsize 2140 & \scriptsize 768
    & \scriptsize 747 & \scriptsize 812\\
\hline
\end{tabular}
}

  \end{center}
\end{table}

\paragraph{Analysis of the number of iterations.}
%
%
In order to validate the abstraction mechanism described in
\autoref{sec:abstract}, we considered the LMHS-nes variant, 
and the benchmarks generated according to the example in
\autoref{fig:graph01}.
The reason to consider LMHS-nes is that soft clauses are all unit and
the set of soft clauses includes the complete set of variables of the
formula. If the option is not set, then the complete CNF formula is
replicated inside the MIP solver (CPLEX), \emph{as a preprocessing
  step}, which results in exactly one call to
CPLEX~\cite{davies-sat13}.

\autoref{tab:maxis-numbers} presents the results obtained, where first
and second row show the $k$ and the $m$ parameters of the instance.
The third row (UB) shows the upper bound on the number of iterations
presented in \autoref{sec:abstract}.
The fourth and fifth rows show the number of disjoint cores (\#DC) and
the number of iterations (\#I) reported by LMHS-nes.
%
As can be concluded from the table, the number of iterations is always 
smaller than the upper bound, suggesting that the algorithm is able to
abstract clauses more effectively than in the worst case scenario.
The ability of \hmxhs algorithms to find good abstractions is expected
to represent a significant step into deploying \hmxsat problem
solvers.


%
%
%

\section{Conclusions \& Research Directions} \label{sec:conc}

The practical success of recent \mxsat solvers not only motivates
investigating novel applications, but it also justifies considering
subclasses of the general \mxsat problem.
This paper investigates the subclass of \mxsat restricted to Horn
clauses, i.e.\ \hmxsat. The paper shows that a comprehensive set of
optimization and decision problems are either formulated as \hmxsat or
admit simple reductions to \hmxsat. The paper also shows that
fundamental decision problems, including SAT and CSP, can be
reduced to \hmxsat.
Although NP-hardness of \hmxsat guarantees that such reductions must
exist, the paper develops simple reductions, some of which were
unknown to our best knowledge.
The paper also proposes a \hmxsat algorithm, based on a well-known
family of \mxsat algorithms~\cite{bacchus-cp11,jarvisalo-sat16}, but
which exploits the fact that the formulas to be analyzed are Horn.
The experimental results show the promise of reductions of \hmxsat and
motivate investigating further the use of \hmxsat as a generic problem
solving approach. This also motivates the development of more efficient
implementations of \hmxhs and of alternative approaches to \hmxhs.



\bibliography{refs}
\bibliographystyle{abbrv}

\end{document}